\documentclass[sigplan,10pt, screen]{acmart}
\renewcommand\footnotetextcopyrightpermission[1]{}

\usepackage{enumitem}
\usepackage{multirow}
\usepackage{makecell}
\usepackage{tcolorbox}
\usepackage{listings}
\usepackage[ruled, linesnumbered, vlined]{algorithm2e}
\usepackage{booktabs}

\begin{document}

\title{\sys{}: Micro-Serving Text-to-Image Diffusion Workflows}

\author{
    \rm{Lingyun Yang}$^{\dag *}$,
    \rm{Suyi Li}$^{\dag * \#}$,
    \rm{Tianyu Feng}$^{\dag}$,
    \rm{Xiaoxiao Jiang}$^{\dag}$,
    \rm{Zhipeng Di},
    \rm{Weiyi~Lu},
    \rm{Kan~Liu},
    \rm{Yinghao~Yu},
    \rm{Tao Lan},
    \rm{Guodong Yang},
    \rm{Lin Qu},
    \rm{Liping Zhang},
    \rm{Wei Wang}$^{\dag}$
    \\
    $^{\dag}$Hong Kong University of Science and Technology
    \quad Alibaba Group
}


\newcommand{\sys}{\textsc{LegoDiffusion}}
\newcommand{\diffusers}{\textsc{Diffusers}}
\newcommand{\clockwork}{\textsc{Diffusers-C}}
\newcommand{\shepherd}{\textsc{Diffusers-S}}

\newcommand{\PHB}[1]{\noindent\textbf{#1}} 
\newcommand{\PHM}[1]{\vspace{.4em} \noindent\textbf{#1}} 

\newcommand{\secref}[1]{\S\ref{#1}}
\newcommand{\figref}[1]{Fig.~\ref{#1}}
\newcommand{\tabref}[1]{Table~\ref{#1}}
\newcommand{\thmref}[1]{Theorem~\ref{#1}}
\newcommand{\prgref}[1]{Program~\ref{#1}}
\newcommand{\algref}[1]{Algorithm~\ref{#1}}
\newcommand{\eqnref}[1]{Equation~\ref{#1}}
\newcommand{\clmref}[1]{Claim~\ref{#1}}
\newcommand{\lemref}[1]{Lemma~\ref{#1}}
\newcommand{\ptyref}[1]{Property~\ref{#1}}

\newcommand{\eg}{{e.g.\@\xspace}}
\newcommand{\ie}{{i.e.\@\xspace}}
\newcommand{\etc}{
        \@ifnextchar{.}
        \textit{etc}
        \textit{etc.\@\xspace}
}

\newcommand{\term}{\textsf}
\newcommand{\code}{\texttt}
\newcommand{\ths}{\textsuperscript{th}}
\newcommand{\circledtext}[1]{\raisebox{.5pt}{\textcircled{\raisebox{-.9pt} {#1}}}}

\newcommand{\todo}[1]{\noindent\textcolor{red}{[TODO: #1]}}
\newcommand{\todowriting}[1]{\noindent\textcolor{red}{[Writing: #1]}}
\newcommand{\todofigure}[1]{\noindent\textcolor{red}{[Figure: #1]}}
\newcommand{\todoexp}[1]{\noindent\textcolor{red}{[Experiment: #1]}}
\newcommand{\suyi}[1]{\noindent\textcolor{violet}{#1}}
\newcommand{\wei}[1]{\textcolor{red}{#1}}
\newcommand{\lingyun}[1]{\textcolor{cyan}{#1}}

\definecolor{mygreen}{rgb}{0,0.6,0}
\definecolor{mygray}{rgb}{0.5,0.5,0.5}
\definecolor{mymauve}{rgb}{0.58,0,0.82}
\definecolor{myred}{rgb}{0.79,0.15,0.15}
\lstdefinestyle{mypython}{
  language=Python,
  backgroundcolor=\color{white},   
  basicstyle=\scriptsize\ttfamily,        
  breakatwhitespace=false,         
  breaklines=true,                 
  captionpos=b,                    
  commentstyle=\color{mygreen},
  deletekeywords={...},            
  escapeinside={\%*}{*)},          
  extendedchars=true,              
  firstnumber=1,                
  frame=no,                    
  keepspaces=true,                 
  keywordstyle=\color{mymauve},       
  language={Python},                 
  morekeywords={DIRECT, PERIODIC, IMMEDIATE, BY_TIME, EVERY_OBJ, deferred},            
  numbers=left,                    
  numbersep=5pt,                   
  numberstyle=\tiny\color{mygray}, 
  rulecolor=\color{black},         
  showspaces=false,                
  showstringspaces=false,          
  showtabs=false,                  
  stepnumber=1,                    
  stringstyle=\color{myred},     
  tabsize=2,                    
  title=\lstname                   
}

\setlength{\abovecaptionskip}{3pt plus 1pt minus 1pt}
\setlength{\belowcaptionskip}{3pt plus 1pt minus 1pt}
\setlength{\abovedisplayskip}{3pt}
\setlength{\belowdisplayskip}{3pt}

\setlength{\skip\footins}{3pt plus 2pt}
\begin{abstract}
Text-to-image generation
executes a \emph{diffusion workflow} comprising multiple models centered on a base diffusion model.
Existing serving systems treat each workflow as an opaque monolith, provisioning, placing, and scaling all constituent models together, 
which obscures internal dataflow, prevents model sharing, and enforces coarse-grained resource management.
In this paper, we make a case for \emph{micro-serving} diffusion workflows with \sys{}, a system that decomposes a workflow into loosely coupled model-execution nodes that can be independently managed and scheduled.
By explicitly managing individual model inference, \sys{} unlocks cluster-scale optimizations, including per-model scaling, 
model sharing, and adaptive model parallelism.
Collectively, \sys{} outperforms existing diffusion workflow serving systems, 
sustaining up to 3$\times$ higher request rates and tolerating up to 8$\times$ higher burst traffic.
\end{abstract}

\maketitle
\pagestyle{plain}

\def\thefootnote{*}\footnotetext{Equal contribution; $^\#$ Corresponding author}

\section{Introduction}
\label{sec:intro}

\begin{figure}[t]
  \centering
  \includegraphics[width=0.99\linewidth]{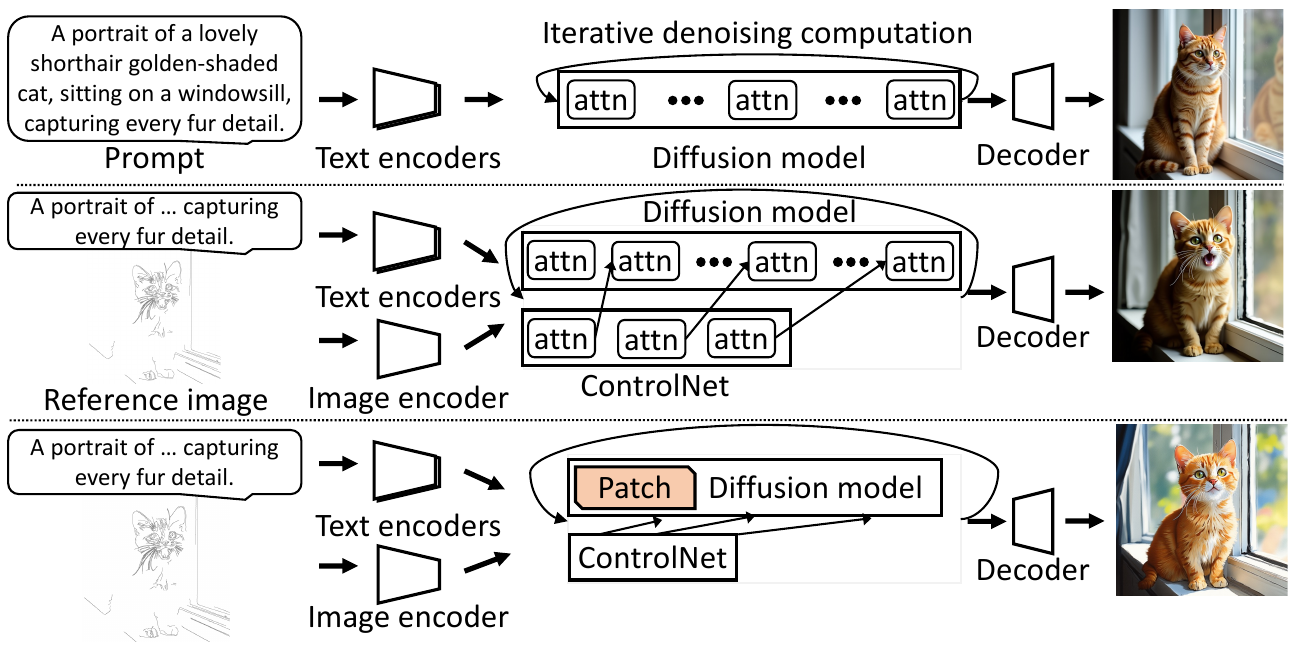}
  \caption{\textbf{Top}: a basic diffusion workflow using Flux-Dev~\cite{flux2024}. \textbf{Middle}: add \textit{ControlNets}~\cite{zhang2023controlnet} alongside the base diffusion model, which process an additional input reference image and pass the intermediaries to diffusion model to control the composition in image generation. 
  \textbf{Bottom}: Further add \textit{LoRA}~\cite{hu2022lora} to change image styles by patching LoRA weights onto diffusion model weights.}
  \label{fig:workflow_example}
\end{figure}

Text-to-image (T2I) generation using diffusion models enables the creation of
high-quality, contextually accurate images from textual description~\cite {ootd,
ju2024BrushNet,zhang2023controlnet, gpt4o,dall-e,
about_openart}. A typical T2I generation \emph{workflow} integrates a \emph
{base diffusion model} with multiple \emph{adapter models} to form a pipeline.
Execution begins with text encoders that convert prompts into
embeddings (\figref {fig:workflow_example}-top). Conditioned on the
embeddings, the diffusion model iteratively generates latent representations via
a denoising process, which are subsequently decoded as the final image.
To refine visual attributes such as composition or artistic styles, T2I
workflows increasingly incorporate adapter models, such as
ControlNet~\cite{zhang2023controlnet} and LoRA~\cite{hu2022lora}, alongside the
base model~\cite{katz,diffusion_production}. By augmenting the diffusion process
(\figref{fig:workflow_example}), these adapters enable fine-grained alignment
with user intent and aesthetics~\cite{zhang2023controlnet,hu2022lora,ye2023IP-Adapter,zhang2025scaling_iclight}.

Despite the modular nature of diffusion workflows, existing inference systems,
such as HuggingFace Diffusers~\cite{diffusers_server} and ComfyUI~\cite
{comfyUI_server}, primarily employ a \emph{monolithic serving} practice, where
the entire workflow, comprising the base model and all associated adapters, is
encapsulated into a monolithic instance for provisioning and scheduling. The
system treats these workflow instances as opaque black boxes running on a fixed
set of GPUs, while remaining oblivious to the workflow's internal model executions and data
exchanges.

While operationally simple, monolithic serving introduces fundamental
inefficiencies. First, it forces \emph{coarse-grained scaling}: if a single
model becomes a bottleneck, the system must replicate the entire workflow.
Second, isolated monoliths \emph{preclude model sharing}. Production
traces~\cite{katz,diffusion_production} show that popular diffusion backbones
and adapters are frequently reused across workflows. However, monolithic serving
forces each workflow instance to maintain its own model copies, leading to
redundant memory footprints. Third, treating workflows as opaque black boxes
hides internal data dependencies, preventing the system from automatically
optimizing resource allocation and pipelining. Finally, tight coupling increases
fragility: a single model failure crashes the entire workflow.

To address these limitations, we argue that the schedulable unit for inference
should be the individual model execution node, not the entire diffusion
workflow. Instead of monolithic instances, the system should decompose workflows
into loosely-coupled microservices---each encapsulating a specific component
like a text encoder, diffusion backbone, or adapter. This \emph{micro-serving}
architecture directly addresses the inefficiencies of monolithic serving. First,
it enables \emph{fine-grained scaling}: individual models can scale elastically
based on real-time demand, eliminating the resource waste of replicating the
entire pipeline. Second, it facilitates \emph{cross-workflow model sharing}:
distinct workflows can multiplex shared models, such as a common base model, avoiding redundant memory footprints. Third, by making model
execution and data flow explicit, the system regains visibility into the
computation graph, enabling \emph{automated optimization} of resource allocation
and pipelining. Decoupling the workflow also enables fault isolation and fast
failure recovery.

However, realizing micro-serving for diffusion workflows presents non-trivial
systems challenges. While prior frameworks have successfully applied
micro-serving to CPU-centric data
analytics~\cite{spark,atoll,caerus,Autothrottle,pheromone} and LLM-based agentic
workflows~\cite{ray,Ayo,langchain,llamaindex}, applying this paradigm to diffusion
pipelines requires overcoming hurdles that these systems cannot handle. First,
diffusion workflows exhibit complex, iterative data dependencies between base
models and adapters, which cannot be expressively defined or easily supported in
existing frameworks. Second, decoupling these tightly integrated model workflows
necessitates massive, latency-sensitive tensor communications across GPUs.
Existing LLM or CPU micro-serving frameworks lack an efficient data plane to
manage these high-bandwidth transfers.

In this paper, we present \sys{}, a system purpose-built for the efficient
micro-serving of diffusion workflow. \sys{} introduces
three key designs:

\PHM{Programming Interface \& Compilation.} \sys{} provides a Python-embedded
domain-specific language (DSL) for composing diffusion workflows. The DSL
exposes primitives for model initialization, inference, and diffusion-specific
operations for LoRA and ControlNet application~\cite{hu2022lora,zhang2023controlnet}. To provide a unified support
for diverse community-developed models~\cite{huggingface_model_downloads},
\sys{} wraps each model behind a standardized interface that encapsulates its
native loading and inference logic. All primitives enforce strict input/output
typing, making data dependencies explicit and catching errors at compile time. A
\emph{graph compiler} translates the workflow composition into a
directed acyclic graph (DAG) of loosely coupled \emph{workflow nodes}. Each node
represents a discrete model inference operator that can be independently
provisioned and scheduled--the fundamental unit of micro-serving.

\PHM{Runtime \& Data Plane.} At runtime, \sys{} applies \emph{lazy execution}:
upon each request, it analyzes the workflow DAG and dynamically recomposes the
compute graph---inserting or substituting nodes---to apply diffusion-specific
optimizations. Decomposing workflows into distributed nodes, however, introduces
high-bandwidth tensor transfers with complex synchronization requirements (e.g.,
forwarding ControlNet intermediates at specific denoising layers). To handle
this, we design a \emph{distributed data engine} atop NVSHMEM~\cite{nvshmem}
that enables GPU-direct, zero-copy tensor movement over high-speed
interconnects. The engine provides two fetch modes---\emph{eager} and
\emph{deferred}---so that tensors arrive precisely when needed without stalling
execution. These mechanisms are transparent to developers: once the compiler
produces the workflow DAG, the data engine automatically orchestrates all
inter-node tensor transfers.

\PHM{Workflow Node Scheduling.} The \sys{} scheduler maps workflow nodes onto
distributed \emph{executors} using three strategies that exploit micro-serving's
decomposition. First, it enforces \emph{model-granular scaling}: rather than
replicating entire workflows, \sys{} scales only the bottleneck models, avoiding
redundant resource provisioning. Second, because a loaded model is
workflow-agnostic, the scheduler preferentially dispatches nodes to executors
that already hold the required model state, enabling \emph{multi-tenant model
sharing}. Third, \sys{} employs \emph{adaptive parallelism}: it dynamically
adjusts model parallelism at scheduling time based on real-time cluster
availability, right-sizing resource allocation to maximize throughput without
incurring queuing delays.

We prototyped \sys{} and evaluated it across a diverse array of diffusion
workflows, encompassing SD3~\cite{sd3}, SD3.5-Large~\cite{sd35}, Flux-Dev, and
Flux-Schnell~\cite{flux2024}, along with their respective adapters. Our
experiments show that \sys{}'s micro-serving architecture significantly
outperforms state-of-the-art monolithic serving systems, sustaining up to
3$\times$ higher request rates and satisfying 6$\times$ more stringent SLO, and
tolerating 8$\times$ higher burst traffic, while meeting latency SLOs for over
90\% of requests. Crucially, we verify that \sys{} maintains full compatibility
with emerging diffusion-specific optimizations, such as approximate
caching~\cite{nirvana}, achieving
performance gains consistent with their original monolithic implementations.
We will open-source \sys{}
 after the double-blind review process.

\section{Background and Problem Statement}
\label{sec:background}

\subsection{A Primer on Image Generation Workflow}
\label{sec:diffusion_primer}

\PHB{Basic Workflows.}
As illustrated in \figref{fig:workflow_example}-top,
a basic text-to-image (T2I) generation workflow consists of three models: a \emph{text encoder}, a \emph{base diffusion model}, and a \emph{decoder-only variational autoencoder} (VAE). 
The process begins with the text encoder, which encodes a text prompt into a sequence of semantic token embeddings. The system then initializes a latent tensor with random Gaussian noise. Conditioned on the text embeddings, the base diffusion model \emph{iteratively refines} this tensor through a series of \emph{denoising steps}. Finally, the denoised latent representation is passed to the VAE decoder, which reconstructs the output image in pixel space.

\PHM{Workflows with Adapters.} To achieve fine-grained control over visual
attributes, such as spatial structure, diffusion models are
frequently augmented with \emph{adapter
models}~\cite{katz,zhang2023controlnet,hu2022lora, zhang2025scaling_iclight,
Fooocus, ye2023IP-Adapter}. These adapters can be categorized into two classes
based on their execution patterns~\cite{katz}:

\textbf{\emph{1) Parallel Execution Adapters.}} The first class includes
adapters that operate \emph{in tandem} with the base diffusion model during
inference, such as ControlNet~\cite{zhang2023controlnet}
(\figref{fig:workflow_example}-middle). 
From a systems perspective, they introduce two complications: (1) their
parameter sizes are often comparable to the base model, introducing substantial
model loading latency; and (2) maximizing throughput often requires
parallelizing the adapter and base model across GPUs, which necessitates
intricate synchronization and data transfer patterns (\S\ref{sec:limitations}).

\textbf{\emph{2) Weight-Patching Adapters.}} The second class adapts the base
model through parameter-efficient fine tuning, such as
LoRA~\cite{hu2022lora} and IC-Light~\cite {zhang2025scaling_iclight}
(\figref{fig:workflow_example}-bottom). These adapters
patch the base model's weights before inference, incurring \emph{no
additional computational overhead} during subsequent denoising steps. The
trade-off is \emph{state management}: once patched, a diffusion model replica is
specialized to a specific request until its weights are restored or replaced.
Serving such workflows therefore requires fetching adapter weights from remote
storage on demand~\cite{katz}, which can bottleneck loading and complicate
sharing model replicas across requests.


\begin{figure}[t]
  \centering
  \includegraphics[width=0.9\linewidth]{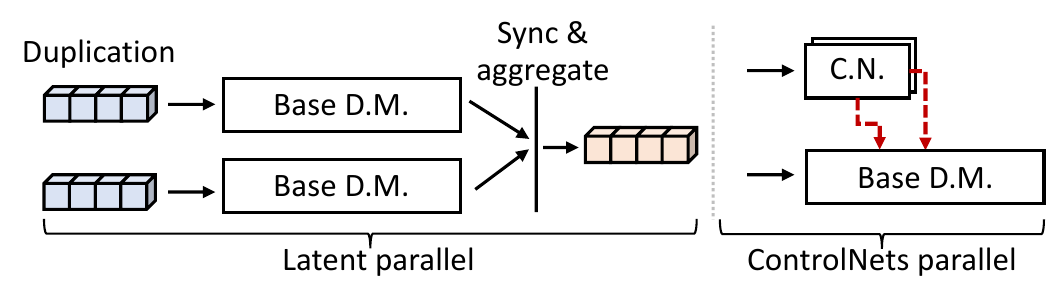}
  \caption{Latent parallelism and ControlNets parallelism.}
  \label{fig:latent_controlnet_parallel}
\end{figure}

\PHM{Data Dependencies in Diffusion Workflows.} Diffusion workflows exhibit
intricate data dependencies induced by performance-oriented parallelization
strategies. These strategies improve
performance~\cite{katz,fang2024xdit,li2024DistriFusion}, but they also introduce
model-specific data transfers and synchronizations that monolithic systems struggle
to express or optimize (\S\ref{sec:system_runtime}). 
We highlight three common cases:

\textbf{\textit{1) Latent Parallelism.}} Diffusion models typically use
classifier-free guidance (CFG)~\cite {ho2021classifierfree} to improve image
quality by executing two denoising passes at each step: one conditioned on the
prompt and one unconditional. Latent parallelism accelerates CFG by
parallelizing these two computations on separate
GPUs~\cite{katz,fang2024xdit,li2024DistriFusion}. However, this approach
introduces frequent ``scatter-gather'' synchronization, where partial results
must be aggregated at \emph{every} denoising step
(\figref{fig:latent_controlnet_parallel}). These high-frequency communication
barriers can erode the benefit of parallelism if not handled efficiently.

\textbf{\textit{2) ControlNet Parallelism.}}
To reduce the overhead of ControlNets, serving systems often execute them in
parallel with the base diffusion model on separate GPUs~\cite{katz}
(\figref{fig:latent_controlnet_parallel}). This design introduces fine-grained
data dependencies: ControlNet feature maps must be transferred to, and consumed
by, specific layers of the base model during each denoising step. The exact
communication pattern depends on the base model and becomes even more complex
when multiple ControlNets are used, producing fan-in/fan-out transfers that
are difficult to schedule efficiently.

\textbf{\textit{3) Asynchronous LoRA loading.}}
In production systems, LoRA adapters are often stored remotely and must be
fetched on demand~\cite{katz}. To hide this fetching cost,
\emph{asynchronous LoRA loading} overlaps adapter retrieval with the early
stages of base-model inference. When the LoRA weights arrive, the system must
pause execution, hot-patch the base model in GPU memory, and then resume
computation~\cite{katz}. This optimization introduces \emph
{non-deterministic timing} and \emph{dynamic state mutation}: execution now
depends on I/O completion, forcing the system to coordinate mid-inference
weight updates without incurring synchronization stalls.

\subsection{Monolithic Serving and Its Inefficiency}
\label{sec:limitations}

Existing diffusion serving systems, such as HuggingFace
Diffusers~\cite{diffusers,diffusers_server}, ComfyUI~\cite {comfyUI},
SGLang-Diffusion~\cite{sglang_diffusion}, vLLM-Omni~\cite{vllm_omni}, and
xDiT~\cite{fang2024xdit}, operate on a \emph{monolithic paradigm}. Whether
employing Diffusers’ ``single-file'' abstraction\footnote{SGLang-Diffusion, vLLM-Omni and xDiT explicitly reuse
the pipeline design from Diffusers~\cite{diffusers_philosophy} to serve diffusion
workflows~\cite{sglang_diffusion, vllm_omni}. However, they currently provide insufficient support
to use adapters in their frameworks~\cite{vllm_omni, sglang_adapter_pr,
sglang_roadmap2025}.}~\cite{diffusers_philosophy,
sglang_diffusion, vllm_omni} or ComfyUI’s flexible node graph~\cite{comfyUI_nodes},
these frameworks encapsulate the entire generation pipeline, comprising the base
model, adapters, and control logic, into a monolithic execution unit.
Consequently, the serving system provisions resources and schedules execution at
the granularity of the entire workflow, without managing internal model
invocations and data flows. 
While this \emph{monolithic serving} simplifies
deployment, it has four
fundamental limitations:

\PHM{L1: Inefficient Scaling via Full Replication.}
Monolithic serving treats the entire workflow as a scaling unit,
enforcing coarse-grained replication regardless of which component is the
actual bottleneck. This \emph{indiscriminate scaling} is particularly costly for
diffusion workloads, where the base diffusion model is typically the sole
bottleneck under load spikes. In standard pipelines~\cite
{flux2024,sd3,podell2024sdxl}, the full workflow footprint is often
1.7$\times$ to 4$\times$ larger than the base model alone. Consequently, scaling the
entire monolith incurs significant overhead: our experiments on NVIDIA H800
GPUs reveal that monolithic replication using Diffusers~\cite
{diffusers} adds up to 80\% in loading latency and wastes up to 75\% of GPU
memory compared to scaling only the bottlenecked component.
Similarly, with vLLM-Omni~\cite{vllm_omni} and SGLang-Diffusion~\cite{sglang_diffusion}, scaling an entire Flux-Dev pipeline adds up to 70\% and 75\% latency, respectively, compared to scaling only Flux-Dev model.

\begin{figure}[t]
  \centering
  \includegraphics[width=0.49\linewidth]{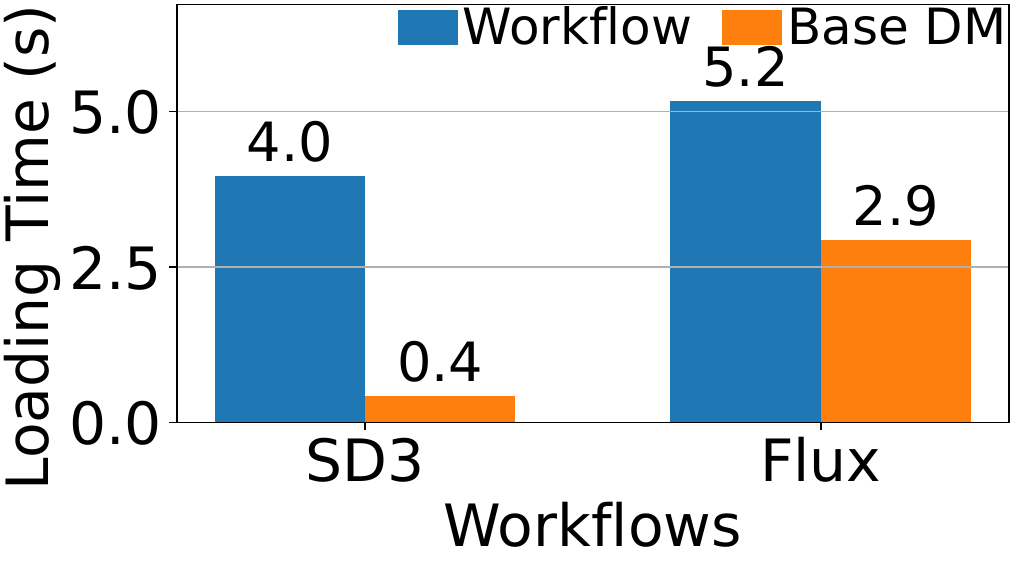}
  \includegraphics[width=0.49\linewidth]{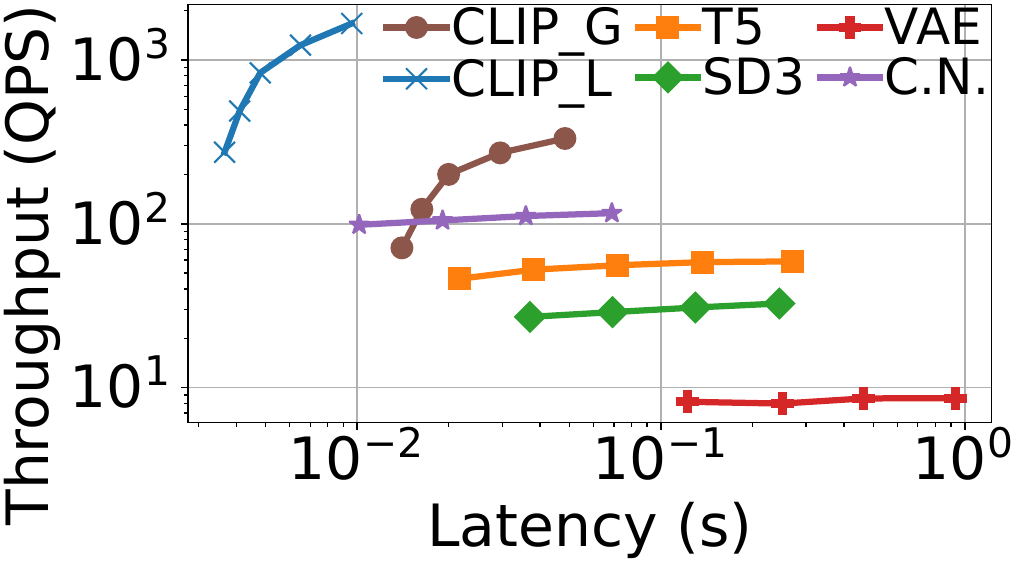}
  \caption{\textbf{Left}: Loading time of workflow scaling and base diffusion model (DM) scaling. \textbf{Right}: Latency-throughput tradeoff of models in a SD3 workflow.
  Both use H800 GPUs.}
  \label{fig:latency_throughput}
  \vspace{-.2in}
\end{figure}

\PHM{L2: Inability to Share Common Models.}
Monolithic serving enforces strict isolation between workflow instances~\cite
{diffusers_server,comfyUI_server}, precluding model sharing. This
design is fundamentally inefficient given that production T2I workloads
exhibit highly skewed model popularity. Alibaba's trace analyses~\cite
{katz,diffusion_production} indicate that popular backbones (e.g., SDXL~\cite
{podell2024sdxl}, SD3~\cite{sd3}, and Flux-Dev~\cite{flux2024}) appear in nearly
all workflows, while the top 5 ControlNets serve
95\% of generation requests. Under monolithic serving, each
workflow instance must maintain independent replicas of these massive models
(2–24 GiB in FP16~\cite{katz}). This redundancy prevents
memory multiplexing, resulting in excessive GPU memory consumption, low GPU
utilization, and load imbalance across replicas.

\PHM{L3: Runtime Inefficiency.}
By encapsulating workflows as opaque black boxes, monolithic serving
eliminates system-level visibility into internal model dependencies, data
flow, and execution logic. This opacity compels the system to enforce rigid,
workflow-level resource allocation, forfeiting opportunities for fine-grained
runtime optimization. Specifically, because models within a workflow exhibit
heterogeneous arithmetic intensities and distinct latency–throughput
trade-offs (\figref{fig:latency_throughput}-right)~\cite{pareto_optimal_throughput}, a static, per-workflow
configuration is inherently suboptimal. Furthermore, monolithic
systems typically enforce a \emph{fixed degree of model parallelism}. Unlike
automatic tuning strategies~\cite{alpaserve,Ayo}, this static approach
prevents the system from adapting to dynamic workloads or fluctuating GPU
availability, leading to significant performance degradation as quantified
in \S\ref{sec:micro_serving_benefits} (\figref
{fig:dy_parallel_model_sharing}-right).

\PHM{L4: System Fragility and Maintenance Overhead.}
Monolithic serving imposes high maintenance overheads by \emph
{violating modular systems principles}. The tight coupling of independent
components creates system fragility, where a failure in a single
sub-component cascades into a complete workflow failure. This lack of fault
isolation complicates debugging, forcing developers to check
the entire monolith to identify root causes. Besides, under the monolithic
architecture, updating a single component
necessitates holistic validation and coordination across the entire workflow,
unnecessarily prolonging development and deployment cycles.

\section{A Case for Micro-Serving}
\label{sec:case_for_micro_serving}

We advocate \emph{micro-serving} diffusion workflows. Instead of treating an
entire workflow as a schedulable unit, micro-serving decomposes the workflow
into independently managed model-execution components and gives the serving
system \emph{per-model} control over scaling, sharing, and runtime configuration. This
design effectively addresses the inefficiencies of current monolithic serving
systems (\S\ref{sec:micro_serving_benefits}). However, it introduces new
challenges for diffusion workflows (\S\ref{sec:challenges}).

\subsection{Benefits of Micro-Serving}
\label{sec:micro_serving_benefits}

\PHB{Per-Model Management.} Micro-serving makes each model, rather than the
entire workflow, the unit of management. This lets the serving system scale
only the bottlenecked model and choose resources according to each model's
latency--throughput tradeoff, directly addressing \textbf{L1} and part of
\textbf{L3}. As a result, the system avoids replicating non-bottleneck
components, reduces model-loading overhead, and better matches heterogeneous
models to available hardware. In \figref{fig:latency_throughput}-left, we
compare full-workflow scaling with scaling only the base diffusion model on
H800 GPUs. Because the diffusion model is the bottleneck, full
replication loads other components unnecessarily. Scaling only the diffusion
model therefore reduces scaling latency by up to 90\%.

\begin{figure}[t]
  \centering
  \includegraphics[width=0.49\linewidth]
  {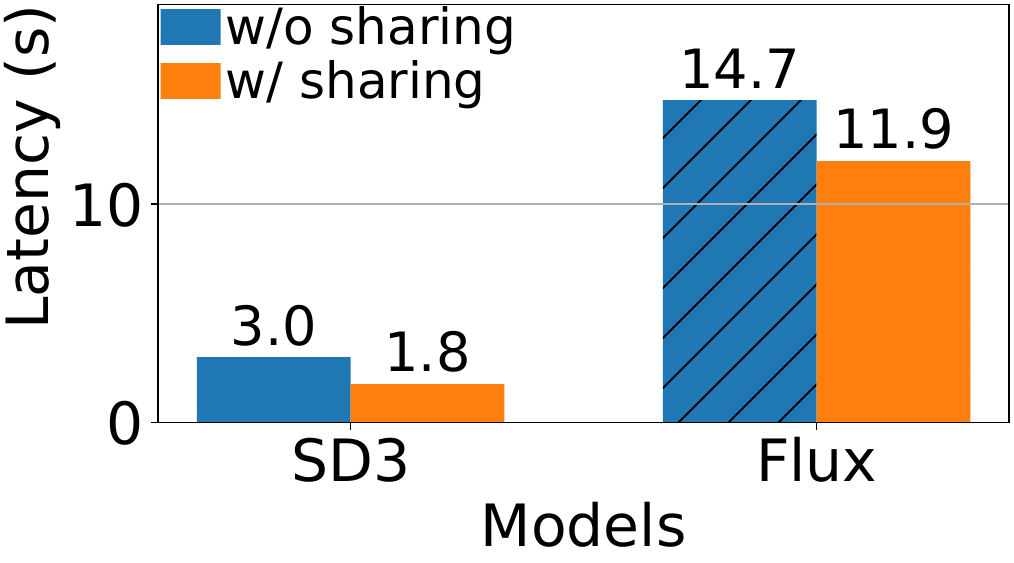}
  \includegraphics[width=0.49\linewidth]{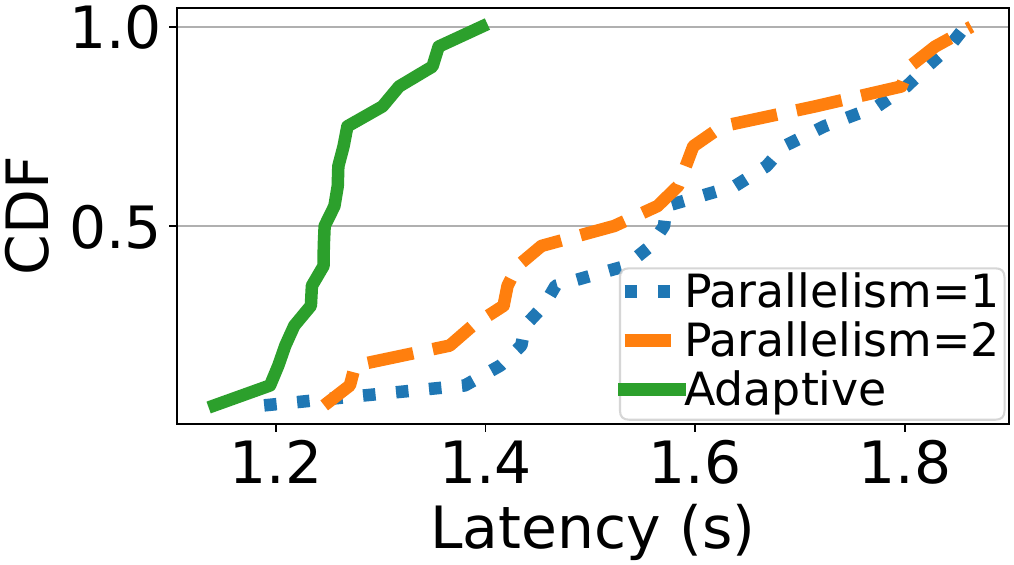}
  \caption{\textbf{Left:} Model sharing reduces request latency. 
  \textbf{Right:} Adaptive parallelization reduces request latency. }
  \label{fig:dy_parallel_model_sharing}
\end{figure}

\PHM{Model Sharing.} When different workflows invoke common models~\cite{katz},
micro-serving lets the system share those loaded replicas across workflows,
directly addressing \textbf{L2}. Instead of binding a model replica to the
workflow that loaded it, the system can multiplex compatible requests onto any
resident replica. This reduces redundant replicas and improves load balance,
since requests can use identical models already loaded elsewhere in the cluster.
To show these benefits, we serve a pair of workflows on two H800 GPUs:
one with ControlNet and one without. This setup creates model-sharing
opportunities for the text encoders and diffusion models. In
\figref{fig:dy_parallel_model_sharing}-left, we compare request latency with
and without model sharing for such workflow pairs, using SD3 and Flux as the
base diffusion model in separate experiments.
Compared with isolated workflow replicas, multiplexing already-loaded models
reduces request latency by up to 40\% and GPU memory footprint by up to 60\%.
Furthermore, base diffusion models patched with adapters (e.g., LoRA, see
\S\ref{sec:diffusion_primer}) can still be shared across requests requiring
different LoRAs through efficient patch swapping~\cite{katz}.
We elaborate on this in \S\ref{sec:microbenchmark}.

\PHM{Adaptive Resource Configuration.}
Micro-serving exposes model dependencies and execution choices to the runtime,
which lets the system tune resource configurations per model instead of fixing
one for the entire workflow. This directly addresses
\textbf{L3}. Automatic model parallelism is one example. We deploy three SD3
workflows~\cite{sd3} on four H800 GPUs under three settings: \emph
{Parallelism=1} fixes the parallelism degree at 1 and leaves acceleration
opportunities unused; \emph{Parallelism=2} always applies \emph{latent
parallel} proposed in~\cite{katz,fang2024xdit}, which speeds up image
generation without quality loss but requires a pair of GPUs
(\S\ref{sec:diffusion_primer}); and \emph{Adaptive} selects the parallelism
degree at runtime according to GPU availability. As shown in
\figref{fig:dy_parallel_model_sharing}-right, the tradeoff is clear:
\emph{Parallelism=1} yields consistently higher latency because it forgoes
parallel speedup, whereas \emph{Parallelism=2} introduces queuing when later
requests wait for an available GPU pair, producing a stepped CDF curve.
Compared with the static configurations, \emph{Adaptive}'s
automatic parallelization tuning accelerates average request
serving by 1.3$\times$ and 1.2$\times$, respectively.

\PHM{Modular Development.}
Micro-serving also improves modularity, directly addressing \textbf{L4}. By
serving workflow models as independent components, it gives developers clearer
failure boundaries and cleaner update paths. They can modify, validate, and
debug one model without reasoning about the entire workflow, and they can test
individual components in isolation. 
This reduces maintenance overhead and
makes
bugs easier to localize.

\subsection{Challenges of Micro-Serving}
\label{sec:challenges}

Despite these benefits, \emph{micro-serving} diffusion workflows cannot be built
by directly reusing existing microservice systems for analytics, general task
runtimes, or LLM agents~\cite
{parrot,Ayo,ray,pheromone,spark,atoll,caerus,Autothrottle,mapreduce}. Diffusion
workflows couple heterogeneous models through iterative denoising, adapter
patching, and diffusion-specific parallelism, creating requirements these
systems do not target.

First, the system needs an abstraction that is expressive for developers yet
structured for backend analysis. Analytics DAG systems such as
Spark~\cite{spark} and generic task runtimes such as Ray~\cite{ray} can
express tasks and dependencies, but not adapter--base-model interactions,
deferred tensor dependencies, or patching operations
(\S\ref{sec:diffusion_primer}).

Second, the system must compile a workflow into executable model-level tasks
while preserving diffusion-specific optimizations. LLM-centric systems such as
Parrot and Ayo~\cite{parrot,Ayo} primarily optimize autoregressive inference
through prefix sharing and streamed decoding~\cite{zheng2024sglang}. Diffusion
workflows instead require a compiler that preserves denoising structure and
supports caching, asynchronous LoRA loading, and specialized multi-GPU
parallelization~\cite
{nirvana,katz,li2024DistriFusion,fang2024xdit}.

Third, the runtime must be GPU-native and diffusion-aware. Data analytics and
microservice systems largely assume CPU execution and host-memory
dataflow~\cite{pheromone,spark,caerus,mapreduce,atoll,Autothrottle}, whereas
diffusion workflows exchange \texttt{CUDA} tensors across GPUs, rely on
asynchronous and interleaved data movement, and require a dedicated data
engine.

Finally, micro-serving only pays off if the scheduler can exploit
diffusion-specific opportunities at runtime. Existing systems do not directly
support model-granular scaling, cross-workflow model sharing, adaptive
parallelization, or SLO-aware admission control. These challenges drive our
design of a diffusion-specific programming interface, graph compiler,
runtime/data engine, and scheduler in \S\ref{sec:design} and
\S\ref{sec:cluster_optimizations}.

\section{System Design of \sys{}}
\label{sec:design}

In this section, we present \sys{}, an efficient serving system for
\emph{micro-serving} diffusion workflows. \sys{} comprises four components:
a programming interface for workflow composition and model integration
(\S\ref{sec:programming_interface}), a graph compiler that decomposes
workflows into executable nodes and applies diffusion-specific optimizations
(\S\ref{sec:compiler_with_optimizations}), a runtime with a GPU-native data
engine for workflow execution (\S\ref{sec:system_runtime}), and an
orchestrator for scheduling and resource management
(\S\ref{sec:cluster_optimizations}). Together, these components optimize
cluster-level serving performance while accommodating existing model
acceleration techniques~\cite{diffusers_accelerations}.

\begin{figure}[t]
  \centering
  \includegraphics[width=0.99\linewidth]{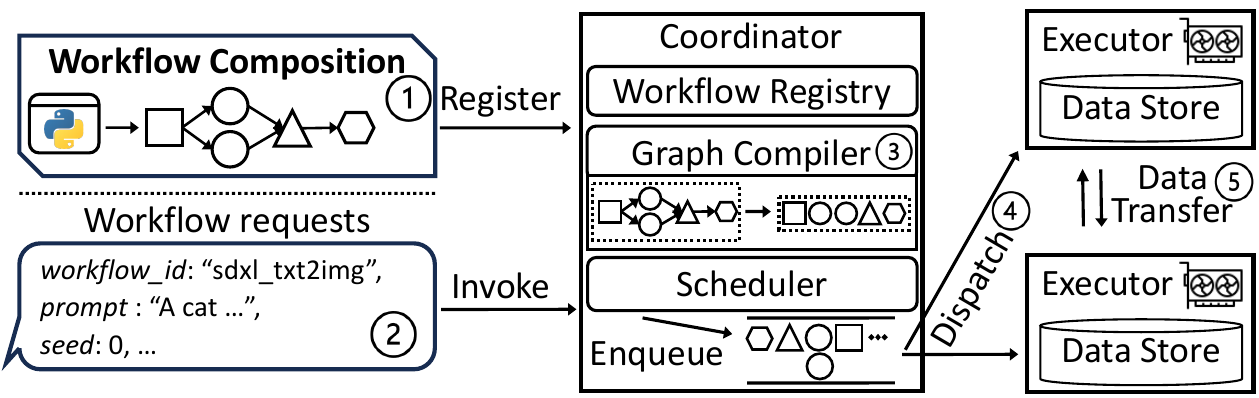}
  \caption{An overview of \sys{}.}
  \label{fig:architecture}
\end{figure}

\PHM{System Overview.} \figref{fig:architecture} presents an overview of \sys{}.
At the frontend, \emph{model developers} integrate individual models by
subclassing a \texttt{Model} base class, and \emph{workflow developers}
compose these models into workflows and register them with the system
(\circledtext{1}). End users invoke registered workflows (\circledtext{2}) by
submitting requests with inputs such as textual prompts and random seeds.

At the backend, the graph compiler transforms a workflow into a set of
loosely coupled workflow nodes (\circledtext{3}), which are dispatched by the
scheduler across a cluster of distributed executors (\circledtext{4}). Each
executor owns one GPU and uses an efficient data engine for inter-node
communication (\circledtext{5}).

\subsection{Programming Model for Developers}
\label{sec:programming_interface}

\sys{}'s programming model targets two audiences.
\emph{Model developers} integrate individual models and adapters by subclassing
a \texttt{Model} base class; they implement model-specific logic without
reasoning about how their models are composed into a workflow. On the other hand,
\emph{workflow developers} assemble models into end-to-end workflows by
instantiating models and invoking them; they do not manually wire a DAG (directed acyclic graph).
Instead, \sys{} adopts an \emph{implicit} workflow programming model: model
invocations and the I/O interfaces declared in each \texttt{Model} subclass
are sufficient for the graph compiler (\S\ref{sec:compiler_with_optimizations})
to infer the workflow DAG and optimize execution automatically.
This contrasts with explicit-DAG systems such as ComfyUI~\cite{comfyUI},
where developers must manually specify every node and edge.

\begin{table}[t]
  \footnotesize
  \centering
  \def\arraystretch{1.1}
  \begin{tabular}{@{}c | l | l @{}}
      \hline
      \textbf{Class}  & \textbf{API}  &  \textbf{Description} \\
      \hline
      \hline

       \multirow{7}{*}{ \texttt{Model} }
          & \texttt{\_\_init\_\_()}   & Create a model instance \\
          \cline{2-3}
          & \texttt{\_\_call\_\_()} & Express a model invocation in the frontend \\
          \cline{2-3}
          & \texttt{setup\_io()} & Define model inputs and outputs \\
          \cline{2-3}
          & \texttt{load()} & Load the model in the backend \\
          \cline{2-3}
          & \texttt{execute()} & Execute model inference in the backend \\
          \cline{2-3}
          & \texttt{add\_patch()} & Attach a patchable adapter to a model \\
          \cline{2-3}
          & \texttt{rm\_patch()} & Remove a patchable adapter from a model \\
      \hline
      \multirow{3}{*}{ \texttt{Workflow}}
          & \texttt{\_\_init\_\_()}   & Create a workflow instance \\
          \cline{2-3}
          & \texttt{add\_input()}   & Add a workflow input placeholder \\
          \cline{2-3}
          & \texttt{add\_output()}   & Add a workflow output placeholder \\
          \cline{2-3}
      \hline
  \end{tabular}
  \caption{Primitives in \sys{}'s Python library for defining models and composing diffusion workflows.}
  \label{tab:primitives}
  \vspace{-.2in}
\end{table}

\PHM{Model Integration.} To keep pace with the proliferation of models and
acceleration techniques~\cite{diffusers_accelerations}, \sys{} provides a
\texttt{Model} base class that standardizes model and adapter integration while
encapsulating all workflow-facing logic.

A model developer subclasses \texttt{Model} and implements three methods
(\tabref{tab:primitives}):
\texttt{setup\_io()} declares the model's typed inputs and outputs, which are
visible to the compiler;
\texttt{load()} initializes the model on a given device; and
\texttt{execute()} runs inference.
Because these are the \emph{only} methods a model developer writes, model
integration is decoupled from workflow construction: a model developer never
reasons about how the model is wired into a larger workflow.

The base class handles workflow integration automatically.
Its \texttt{\_\_call\_\_()} method records each model invocation as a workflow
node and derives data dependencies from the I/O interface declared in
\texttt{setup\_io()}.
This separation captures a key design principle: model developers specify what
a model consumes and produces; \sys{} uses that specification to place the
model into an inferred workflow graph.

\begin{figure}[tb]
\begin{lstlisting}[style=mypython, escapechar=|]
## No need to modify it, invisible to model developers ##
class Model:
  def __init__(self, **kwargs):
    self.setup_io()
    # store associated weight-patching adapters
    self._patches = []
    
  # Make the class callable to create workflow node
  def __call__(self, **kwargs):
     workflow = WorkflowContext.get_current_workflow()
     workflow_node = WorkflowNode(op=self, **kwargs)
     workflow.add_workflow_node(workflow_node)
     return workflow_node.get_outputs()

  @abstractmethod
  def setup_io(self):
    pass

  def add_patch(self, patch):
    self._patches.append(patch)

  def rm_patch(self, patch):
    self._patches.remove(patch)

## Model developers start here ##
class Flux(Model):
  def setup_io(self):
    # define inputs
    self.add_input("latents", torch.Tensor)
    self.add_input("prompt_embeds", torch.Tensor)
    # define "deferred" inputs, detailed in Sec. 4.3.2
    self.add_input("controlnet_inputs", torch.Tensor, deferred=True)   
    # define outputs
    self.add_output("noise_pred", torch.Tensor)
    
  def load(self, model_path, device):
    model = SD3Transformer2DModel.from_pretrained(
        ...
    ).to(device)
    return {"transformer": model}

  @torch.no_grad()
  def execute(self, model_components, **kwargs):
    transformer = model_components["transformer"]
    noise_pred = transformer(**kwargs)[0]
    return {"noise_pred": noise_pred}
\end{lstlisting}
\vspace{-.3in}
\caption{A simplified case of integrating Flux with \sys{}.}
\label{fig:flux_model_integration}
\vspace{-.2in}
\end{figure}


\figref{fig:flux_model_integration} illustrates this split with a simplified
Flux integration: the base \texttt{Model} class (top) is provided by the
framework, while the \texttt{Flux} subclass (bottom) contains only
model-specific code.

\PHM{Workflow Composition.}
Workflow developers compose workflows \emph{declaratively}: they declare
workflow inputs and outputs, instantiate models, and invoke them.
They never explicitly wire a DAG; the graph compiler
(\S\ref{sec:compiler_with_optimizations}) infers all data dependencies from
model invocations and the I/O interfaces declared in \texttt{setup\_io()},
then optimizes the resulting graph.

\figref{fig:workflow_definition} shows a workflow for the Flux example in
\figref{fig:workflow_example}-bottom. Creating a \texttt{Workflow} instance
(line 2) establishes a scope (maintained by \texttt{WorkflowContext}); subsequent
model calls within that scope are automatically recorded as workflow nodes.
Workflow inputs are declared with \texttt{add\_input()}, and intermediate values
such as \texttt{prompt\_embeds} flow directly between model calls. The loop
(line 23) shows that iterative denoising is expressed naturally in Python, while
\sys{} captures the structure needed for backend execution. After construction,
the workflow developer registers the workflow with \sys{} for later invocation
by end users.

\begin{figure}[tb]
\begin{lstlisting}[style=mypython, escapechar=|]
# create a workflow instance
workflow = Workflow(name="flux_txt2img_workflow")
# initialize models. All inherit the Model class
latents_generator = LatentsGenerator()
text_enc = FluxTextEncoder(model_path=model_path)
flux = Flux(model_path=model_path)
controlnet = ControlNet(model_path=controlnet_path)
lora = LoRA(model_path=lora_path)
vae = FluxVAE(model_path=model_path)
# initialize input placeholders for the workflow
seed = workflow.add_input(name="seed", data_type=int)
prompt = workflow.add_input(name="prompt", data_type=str)
num_denoising_steps = workflow.add_input(name="num_denoising_steps", data_type=int)
ref_image = workflow.add_input(name="ref_image", data_type=Image)
# add_patch() registers a LoRA adapter with the flux, 
flux.add_patch(lora)
# Invoke models and establish their I/O dependencies.
# Model invocation is expressed via __call__(). 
latents = latents_generator(seed)
prompt_embeds = text_enc(prompt)
ref_image = vae(image=ref_image, mode="encode")
# Perform iterative denoising computation.
for i in range(num_denoising_steps):
    controlnet_outputs = controlnet(latents, prompt_embeds, ...)
    noise_pred = flux(latents, prompt_embeds, controlnet_outputs, ...)
    latents = denoise(noise_pred, latents) 
output_img = vae(latents, mode="decode")
workflow.add_output(output_img, name="output_img")
\end{lstlisting}
\vspace{-.3in}
\caption{A simplified diffusion workflow using Flux~\cite{flux2024}.}
\label{fig:workflow_definition}
\vspace{-.2in}
\end{figure}

\subsection{Graph Compiler}
\label{sec:compiler_with_optimizations}

The graph compiler (\circledtext{3}) lowers a registered workflow into a
topologically sorted DAG of schedulable workflow nodes and applies optimization
passes before execution.

\PHM{DAG Construction.}
As described in \S\ref{sec:programming_interface}, each model invocation during
workflow composition is recorded as a workflow node with typed I/O declared in
\texttt{setup\_io()}. The compiler resolves data dependencies among these nodes
and produces a topologically sorted DAG. Topological order guarantees correct
execution and exposes optimization opportunities: nodes without mutual
dependencies can run in parallel to reduce latency, and nodes that invoke the
same model can be batched to improve throughput
(\S\ref{sec:system_runtime}).

\PHM{Optimization Passes.}
After constructing the DAG, the compiler applies a series of graph-rewriting
passes. Each pass pattern-matches on node properties (e.g., model type, adapter
attachments) and may insert, remove, or replace nodes. This design makes the
compiler extensible: adding a new optimization requires only a new pass, without
modifying the core lowering logic. The compiler also applies per-model
optimizations such as \texttt{torch.compile()} within individual nodes. Below,
we illustrate two diffusion-specific passes and evaluate them in
\S\ref{sec:eval_case_study}.

\textbf{\emph{1) Approximate caching}}~\cite{nirvana} reduces the number of
denoising steps by initializing from a pre-cached image of a similar prompt
instead of random noise (\S\ref{sec:diffusion_primer}). When a prompt cache is
configured, the compiler replaces the random-latent-initialization node with a
cache-lookup node, requiring no changes to the workflow definition.

\textbf{\emph{2) Asynchronous LoRA loading}}~\cite{katz} overlaps LoRA adapter
retrieval with the early stages of diffusion model inference
(\S\ref{sec:diffusion_primer}). When the compiler detects an
\texttt{add\_patch()} attachment on a model, it rewrites the workflow graph by
inserting (1) an initial node that triggers asynchronous LoRA loading and (2) a
check node after each diffusion-model node that tests whether the adapter is
ready to be patched in. The workflow developer writes only
\texttt{add\_patch(lora)}; the compiler can insert the asynchronous-loading
machinery automatically.

\subsection{\sys{}'s Runtime}
\label{sec:system_runtime}

\subsubsection{Micro-Serving Control Plane}
\label{sec:control_plane}

Given the topologically sorted DAG from the compiler, the runtime executes each
request through a node-level control plane. Under \emph{micro-serving}, each
workflow node is independently schedulable on any executor once its
non-deferred inputs are satisfied; the runtime can also configure parallelism
and resource allocation per node based on the node's encapsulated model and
available hardware.

\PHM{Request Execution Lifecycle.}
Workflows are compiled once at registration time; the compiled DAG is
instantiated only when a request arrives~\cite{spark, lazy_eval_wiki,
lazy_evaluation_design_pattern, tensorflow} (\circledtext{2} in
\figref{fig:architecture}). The control plane enqueues all root nodes (those
with no upstream dependencies) and enters a dispatch loop. In each cycle, the
scheduler selects ready nodes and dispatches them to executors
(\circledtext{4}). When an executor completes a node, it reports the result to
the control plane, which marks downstream nodes whose inputs are now satisfied
as ready. This loop continues until all nodes complete and the workflow output
is returned to the end user. The scheduler's placement and batching policies
are detailed in \S\ref{sec:cluster_optimizations}.

\subsubsection{Distributed Data Engine}
\label{sec:data_engine}

Micro-serving introduces frequent data movement between nodes.
In typical diffusion workflows such as SD3~\cite{sd3} and
Flux models~\cite{flux2024}, CUDA tensors account for over 99\% of transferred data
(see Fig.~\ref{fig:communication_tensor_size}-right); for example, an SDXL
workflow with a single ControlNet transfers 5.3\,GiB~\cite{katz}.
Host-memory staging through PCIe is prohibitively slow at this scale.

To avoid CPU staging, \sys{} deploys a distributed data engine with a
per-executor local data store (\figref{fig:architecture}).
The stores are built on NVSHMEM~\cite{nvshmem}, which provides one-sided GPU
communication over NVLink and RDMA, enabling zero-copy sharing within an
executor and high-speed transfers across executors.

\PHM{Data Fetch Modes.}
Diffusion workflows exhibit diverse data-movement patterns due to their
parallelization strategies (\S\ref{sec:diffusion_primer}). \sys{} supports two
fetch modes: \emph{eager}, where an input must be ready before a node begins
execution, and \emph{deferred}, where a node starts execution and fetches the
input at the point of consumption.

\begin{figure}[t]
  \centering
  \includegraphics[width=0.95\linewidth]
  {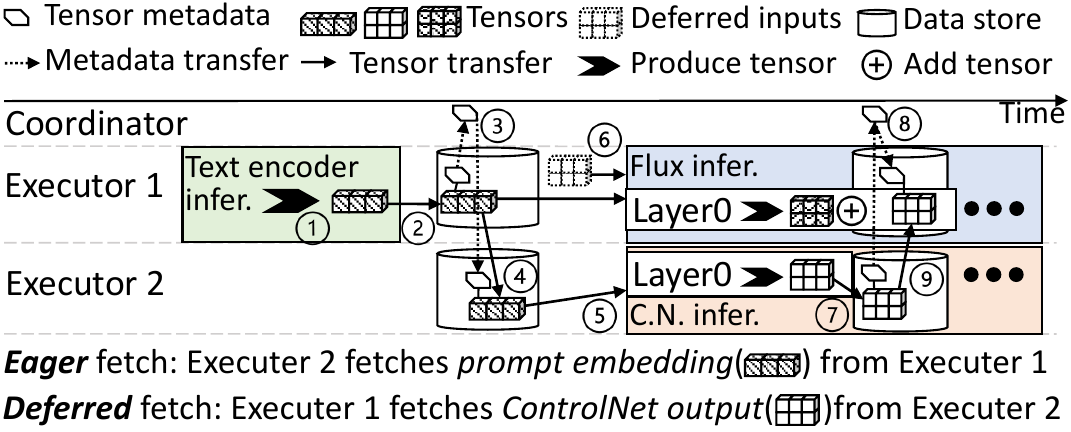}
  \caption{Illustrating data fetch. For simplicity, we primarily illustrate tensor fetch process in data store. \textbf{C.N.}: ControlNet. }
  \label{fig:data_fetch}
  \vspace{-.2in}
\end{figure}

\textbf{\emph{1) Eager data fetch.}}
By default, inputs are fetched eagerly: a node cannot begin until all its eager
inputs are available. In \figref{fig:data_fetch}, the text-encoder node on
Executor~1 produces a prompt embedding (\circledtext{1}) and places it in its
local data store (\circledtext{2}). When the coordinator schedules the
downstream ControlNet node on Executor~2, it forwards the embedding's metadata
(\circledtext{3}). Executor~2 uses this metadata to fetch the tensor into its
own store (\circledtext{4}). The ControlNet node then reads the embedding from
its local store (\circledtext{5}) and begins execution---only after the fetch
completes, a guarantee enforced by eager fetching.

\textbf{\emph{2) Deferred data fetch.}}
Model developers can mark an input as \emph{deferred} (line 32 in
\figref{fig:flux_model_integration}), indicating that it needs not be ready when
a node starts inference. A deferred input is implemented as a fetch function
invoked at the point of consumption: it returns immediately if the data is
available, or blocks until the data arrives.

This mode is tailored to diffusion workflows where ControlNet computation is
interleaved with the base model
(\figref{fig:workflow_example}-middle). At runtime
(\figref{fig:data_fetch}), the ControlNet output consumed mid-way through Flux
inference is marked as a deferred input (dashed tensor, \circledtext{6}). Flux
begins execution without it. When Flux reaches the consumption point, the
ControlNet output has been produced and placed in Executor~2's store
(\circledtext{7}); its metadata is forwarded to Executor~1
(\circledtext{8}), which fetches the tensor into its local store
(\circledtext{9}) for Flux to consume. Without deferred fetching, Flux could
not start until ControlNet completes, eliminating the parallelism between the
two models.

Note that tensor metadata, including a tensor's pointer, is tiny (on the order
of KiB). Executors can piggyback it on node-completion notifications, allowing
the coordinator to track global tensor placements with little overhead.

%

\PHM{Design Properties.}
The data engine is transparent to both model developers and workflow developers:
once the compiler lowers a workflow into nodes, the engine automatically
orchestrates all inter-node data movement.
All intermediate data is \emph{immutable}---tensors produced during diffusion workflow
execution are consumed once and never
updated~\cite{katz,diffusers}---which obviates consistency protocols and
simplifies fault tolerance.
The engine reclaims tensors as soon as no downstream node requires them,
reducing memory pressure. If an executor fails, \sys{} reconstructs lost data
by re-executing the affected nodes, following a similar approach to prior
cluster computing frameworks~\cite{ray, spark, pheromone}.

\section{Workflow Node Scheduling}
\label{sec:cluster_optimizations}

The scheduler is the component that translates \emph{micro-serving} into runtime
actions (\circledtext{4} in \figref{fig:architecture}). It sits between the
compiled workflow DAGs and the executor cluster, maintaining a global queue of
workflow nodes and making three online decisions in each scheduling
cycle: (1)~\emph{which same-model nodes to batch together and which executors to route
them to, exploiting model sharing across workflows}; (2)~\emph{how many GPUs to
allocate per batch, adapting parallelism to current resource availability}; and
(3)~\emph{whether to admit or reject incoming requests to preserve SLO attainment}.
\algref{alg:scheduling} summarizes the scheduling loop; the remainder of this
section describes how the scheduler makes each decision.

To make these decisions, the scheduler maintains two key data structures that
the runtime keeps up to date. A \emph{model state table} records, for every
executor, which models are currently loaded in GPU memory. Executors piggyback
their model states on node-completion notifications to the coordinator, so the
table is updated without extra RPCs. A set of \emph{per-model latency
profiles}, collected
offline, provides stable estimates~\cite{xia2025tridentservestagelevelservingdiffusion} of data-fetch time, model-loading time, and
inference time for each model under various batch sizes and parallelism degrees.
Following prior work~\cite{zhang2023SHEPHERD, alpaserve, toppings, mooncake,
chen2024Punica}, the scheduler orders the ready queue by first-come-first-serve
(FCFS). 
For nodes with the same arrival time (\eg, nodes from the same request), it further prioritizes those at shallower depths in the DAG.
Since optimal ordering requires foreknowledge of future
arrivals~\cite{zhang2023SHEPHERD,gujarati2020Clockwork}, FCFS is a simple,
neutral baseline that isolates \sys{}'s gains from scheduling policy. The
scheduler is a pluggable module; other policies can be substituted.

\begin{algorithm}[ht]
\caption{Scheduling Algorithm}
\label{alg:scheduling}
\footnotesize
\SetKwProg{Fn}{Function}{:}{}

\While{True}{
    \tcp{Admission control runs asynchronously (\S\ref{sec:sched_admission})}
    Admit or reject arrived requests\;
    \tcp{Identify ready nodes}
    $Q_{ready} \leftarrow$ nodes with satisfied dependencies from queue\;

    $E_{avail} \leftarrow$ currently available executors\;

    \If{$Q_{ready} = \emptyset$ \textbf{or} $E_{avail} = \emptyset$}{
        \textbf{continue}\;
    }

    Sort $Q_{ready}$ by (arrival time, node depth)\;
    $n_{head} \leftarrow Q_{ready}.pop(0)$\;

    \tcp{Batch same-model nodes (\S\ref{sec:sched_batching})}
    $B_{max} \leftarrow$ profiled max batch size for $n_{head}.model$\;
    $Batch \leftarrow \{n_{head}\} \cup \{n' \in Q_{ready} \mid n'.model = n_{head}.model \text{ and } |Batch| < B_{max}\}$\;

    \tcp{Choose parallelism degree (\S\ref{sec:sched_parallelism})}
    $k_{max} \leftarrow$ max useful parallelism for $n_{head}.model$\;
    $k \leftarrow \min(|E_{avail}|,\; k_{max})$\;

    \tcp{Score and select executors}
    \For{$e \in E_{avail}$}{
        $L_{data} \leftarrow$ CalcDataFetchLatency($Batch$, $e$)\;
        $L_{load} \leftarrow$ $e$ hosts $n_{head}.model$ ? $0$ : CalcLoadTime($n_{head}.model$)\;
        $L_{infer} \leftarrow$ CalcInferenceTime($Batch$, $e$, $k$)\;
        $e.score \leftarrow L_{data} + L_{load} + L_{infer}$\;
    }

    $E_{target} \leftarrow$ top $k$ from $E_{avail}$ with min scores\;

    Dispatch $Batch$ to $E_{target}$ (triggers model load if needed)\;
}
\end{algorithm}

\subsection{Cross-Workflow Batching and Model Sharing}
\label{sec:sched_batching}

In each scheduling cycle, the scheduler pops the FCFS-earliest node $n_{head}$
from the ready queue and inspects its \texttt{model} field. It then scans the
remaining ready nodes for any that reference the \emph{same} model, regardless
of the originating workflow, and groups them into a single batch of up to
$B_{max}$ entries. The per-model $B_{max}$ is determined offline by profiling
batching efficiency: beyond a model-specific threshold, larger batches increase
latency with diminishing throughput gain~\cite{chen2024Punica}. Because matching
is by model identity rather than by workflow, a single batch may contain nodes
from multiple workflows---this is how the scheduler realizes \emph{model
sharing} (\S\ref{sec:micro_serving_benefits}).

After forming a batch, the scheduler must select an executor (Line 13 --17). For each candidate
executor $e$, it computes a latency score from three profiled components:
(1)~$L_{data}$, the cost of fetching the batch's input tensors from their
producing executors via the data engine (\S\ref{sec:data_engine});
(2)~$L_{load}$, the cost of loading the required model into GPU memory; and
(3)~$L_{infer}$, the estimated inference time for the batch. The model state
table makes $L_{load}$ zero for any executor that already hosts the required
model, so the scoring function naturally routes batches to executors with warm
models. When no executor hosts the model, the scheduler selects the executor
with the lowest total score and triggers a model load---loading only the single
needed model, not the entire workflow, unlike monolithic scaling (\textbf{L1} in
\S\ref{sec:limitations}). We evaluate the throughput and latency gains from
model sharing in \S\ref{sec:microbenchmark}.

\subsection{Adaptive Parallelism}
\label{sec:sched_parallelism}

\sys{} exploits two forms of diffusion-specific parallelism
(\S\ref{sec:diffusion_primer}) through scheduling decisions.

\PHM{Inter-Node Parallelism.} When the compiler produces a DAG in which two
nodes have no eager data dependency---for example, a ControlNet and its corresponding
base-model node connected only by a \emph{deferred} input
(\S\ref{sec:data_engine})---both nodes enter $Q_{ready}$ simultaneously once
their non-deferred inputs are satisfied. The scheduler dispatches them to
separate executors in the same or adjacent loop iterations, so they execute
concurrently. Runtime data exchange between the two nodes is handled by \emph{deferred
data fetch}: the base model begins execution immediately, and retrieves the
ControlNet output mid-inference when it becomes available
(\figref{fig:latent_controlnet_parallel}-right).

\PHM{Intra-Node Parallelism.} A single model invocation can also be split across
multiple GPUs via latent parallelism~\cite{katz, li2024DistriFusion,
fang2024xdit}, which partitions the input tensor and distributes the shards to
$k$ executors for parallel inference
(\figref{fig:latent_controlnet_parallel}-left). The scheduler chooses the
parallelism degree $k$ per batch by a \emph{work-conserving} heuristic: it sets $k =
\min(|E_{avail}|,\; k_{max})$, where $k_{max}$ is the maximum useful parallelism
for the model (determined offline). This rule uses all currently available GPUs
without waiting for more to free up, maximizing parallelism while avoiding extra
queueing delay~\cite{alpaserve,Ayo}. Because the scheduler makes this decision
per batch, different invocations of the same model can run at different
parallelism degrees depending on instantaneous cluster load. The scheduler then
selects the $k$ lowest-scoring executors, dispatches the batch with a
parallelism descriptor, and each executor processes its assigned input shard. We
evaluate intra- and inter-node parallelism in \S\ref{sec:microbenchmark}.

\subsection{SLO-Aware Admission Control}
\label{sec:sched_admission}

Admitting requests beyond system capacity inflates queueing delays and causes
cascading SLO violations. \sys{} prevents this with an \emph{early-abort}
admission policy that leverages micro-serving's per-node visibility into request
progress.

When a new request arrives, the scheduler estimates its end-to-end completion
time. Because the control plane tracks which nodes of each inflight request have
completed, the scheduler can compute, for every inflight request, the sum of
profiled latencies along its remaining critical path. It admits the new request
only if the estimated completion time---accounting for current queueing
depth---satisfies the request's latency SLO. Otherwise, the request is rejected
immediately, preserving resources for already-admitted requests. This
early-abort policy is feasible only under micro-serving: in monolithic serving,
the system has no visibility into sub-workflow progress and cannot estimate
remaining work at fine granularity. The admission control runs asynchronously
and does not block the scheduling loop. We quantify its effect on SLO attainment
in \S\ref{sec:microbenchmark}.

\section{Implementation}
\label{sec:implementation}

We have implemented \sys{} with a FastAPI~\cite{fastapi} frontend
and a distributed GPU-based inference engine. The frontend (approx. 1,000 LoC in
Python) exposes an intuitive programming interface for users to compose and
register diffusion workflows (\S\ref{sec:programming_interface}). Users can
invoke registered workflows with customized image generation parameters, such as
prompts and reference images, similar to the OpenAI API~\cite{openai_api}. We
currently support diffusion workflows for popular models including the SD3
family~\cite{sd3} and Flux family~\cite{flux2024}. \sys{}'s backend runtime
consists of a coordinator and distributed executors (\figref{fig:architecture}),
totaling 4,000 lines of Python code. The data engine is implemented in 1,000
lines of C++/CUDA code using NVSHMEM~\cite{nvshmem}. Aside from CUDA tensors,
communication between the coordinator and distributed executors is facilitated
via ZeroMQ~\cite{zmq}.

\section{Evaluation}
\label{sec:evaluation}

We evaluate \sys{} with the following highlights:

\begin{itemize}[topsep=3pt, leftmargin=*, noitemsep, nolistsep, parsep=0pt, partopsep=0pt]
  \item \sys{} outperforms state-of-the-art baselines, sustaining
  up to 3$\times$ higher request rates, satisfying 6$\times$ more stringent SLOs,
  reducing GPU requirements by up to 3$\times$, and tolerating 8$\times$
  higher burst traffic, all while maintaining over 90\% SLO attainment
  (\S\ref{sec:e2e_performance}).

  \item Microbenchmarks isolate the contribution of each scheduling mechanism
  and validate compatibility with emerging diffusion optimizations
  (\S\ref{sec:microbenchmark}, \S\ref{sec:eval_case_study}).

  \item \sys{} introduces negligible system overhead (\S\ref{sec:system_overhead}).
\end{itemize}

\subsection{Experimental Setup}
\label{sec:eval_setup}

\PHB{Diffusion Workflows and Testbed.} We use 12 diffusion workflows composed
from four popular base models: SD3~\cite{sd3}, SD3.5-Large~\cite{sd35},
Flux-Dev~\cite{flux2024}, and Flux-Schnell~\cite{flux2024}. They exhibit diverse
computational characteristics, with parameter counts spanning 2.5B to 12B and
denoising steps ranging from 4 to 50. In~\tabref{tab:eval_settings}, we
categorize these workflows into six evaluation settings (S1--S6) to assess
system performance under varying degrees of workload heterogeneity. We
use a real testbed of 8 to 32 NVIDIA H800 GPUs to evaluate performance and a 256-GPU simulator to
analyze scalability.

\begin{table}[t]
  \centering
  \footnotesize
  \def\arraystretch{0.85} 
  \caption{\textbf{Evaluation Settings}. We use workflows of representative diffusion models.
  S1--S4 represent single-model deployments, each including three workflow variants: a \textit{Basic} workflow (text encoders, diffusion model, and decoder), plus two using adapters (\textit{Basic} + \textit{C.N. 1} and \textit{Basic} + \textit{C.N. 2}). S5--S6 represent mixed-model deployments. \textbf{C.N.}: ControlNet.}
  \label{tab:eval_settings}
  \begin{tabular}{clc}
    \toprule
    \textbf{Setting} & \textbf{Diffusion Model} & \textbf{Workflow}\\
    \midrule
    \multicolumn{3}{l}{\textit{Single-model Deployments (3 workflows each)}} \\
    \cmidrule(l){1-3}
    S1 & SD3~\cite{sd3} & (Basic, +C.N. 1, +C.N. 2) \\
    S2 & SD3.5-Large~\cite{sd35} & (Basic, +C.N. 1, +C.N. 2)  \\
    S3 & Flux-Schnell~\cite{flux2024} & (Basic, +C.N. 1, +C.N. 2)  \\
    S4 & Flux-Dev~\cite{flux2024} & (Basic, +C.N. 1, +C.N. 2) \\
    
    \midrule
    \multicolumn{3}{l}{\textit{Mixed-model Deployments (6 workflows each)}} \\
    \cmidrule(l){1-3}
    S5 & SD3 + SD3.5-Large & S1's + S2's \\
    S6 & Flux-Schnell + Flux-Dev & S3's + S4's \\
    \bottomrule
  \end{tabular}
  \vspace{-.2in}
\end{table}

\PHM{Baselines.} We primarily compare \sys{} with \diffusers{}, the most representative
\emph{monolithic-serving} system (\S\ref{sec:limitations})\cite{diffusers,
diffusers_server}, as our goal is to compare \emph{micro-serving} with the prevailing monolithic design.
While other
systems~\cite{vllm_omni, sglang_diffusion, fang2024xdit} support more parallelism methods and high-performance kernels, they largely inherit
\diffusers{}'s \emph{monolithic} pipeline design and these optimizations are orthogonal to \sys{}.

To compare against a broader monolithic design space, we include \emph{monolithic-serving} system variants by adopting techniques from multi-model serving systems~\cite{gujarati2020Clockwork, zhang2023SHEPHERD} for workflow orchestration. For a fair comparison, the baselines use FCFS scheduling and workflow-level admission control.

\begin{itemize}[topsep=3pt, leftmargin=*, noitemsep, nolistsep, parsep=2pt, partopsep=0pt]

  \item \diffusers{} represents a \emph{static deployment} strategy.
Each workflow is executed \emph{monolithically} and statically bound to dedicated GPUs~\cite{diffusers, diffusers_server}.
It cannot share models across workflows, adapt parallelism at runtime.

  \item \clockwork{} implements a \emph{swap-based} serving strategy by adapting
Clockwork~\cite{gujarati2020Clockwork} to \diffusers{}. Leveraging the
predictable end-to-end latency of diffusion workflows, it treats each
\emph{monolithic} workflow as a swappable DNN model unit, dynamically loading and unloading workflows into GPU memory on demand.
Because the swap unit is an entire workflow, it cannot share individual models across workflows or adapt parallelism within a workflow.

  \item \shepherd{} incorporates the \emph{planning-and-scheduling} framework of
Shepherd~\cite{zhang2023SHEPHERD} to orchestrate instances, modeling each \emph{monolithic} diffusion workflow as a distinct model unit.
Like \clockwork{}, it schedules whole workflows and thus cannot exploit per-model sharing or adaptive parallelism.
\end{itemize}

\PHM{Workloads.} We use a real-world T2I production trace~\cite{katz}. To
rigorously evaluate under diverse conditions, we vary request
arrival rates, SLO targets, traffic burstiness, and testbed sizes, effectively
simulating a wide spectrum of real-world traffic patterns and performance
requirements.

\PHM{Metrics.}
Our primary metric is \textit{SLO attainment}: the fraction of
requests completed within their specified latency
deadline. We set the default deadline to $2\times$ the solo inference latency of each workflow (SLO Scale $= 2$), which is tight given that any queueing or resource contention will cause violations.
Unlike prior works~\cite{katz, li2024DistriFusion, nirvana, fang2024xdit},
\sys{} does not alter the computation performed during diffusion inference and
therefore requires no evaluation of the image quality.

\subsection{End-to-End Performance}
\label{sec:e2e_performance}

\begin{figure*}[t]
  \centering
  \includegraphics[width=1.0\linewidth]{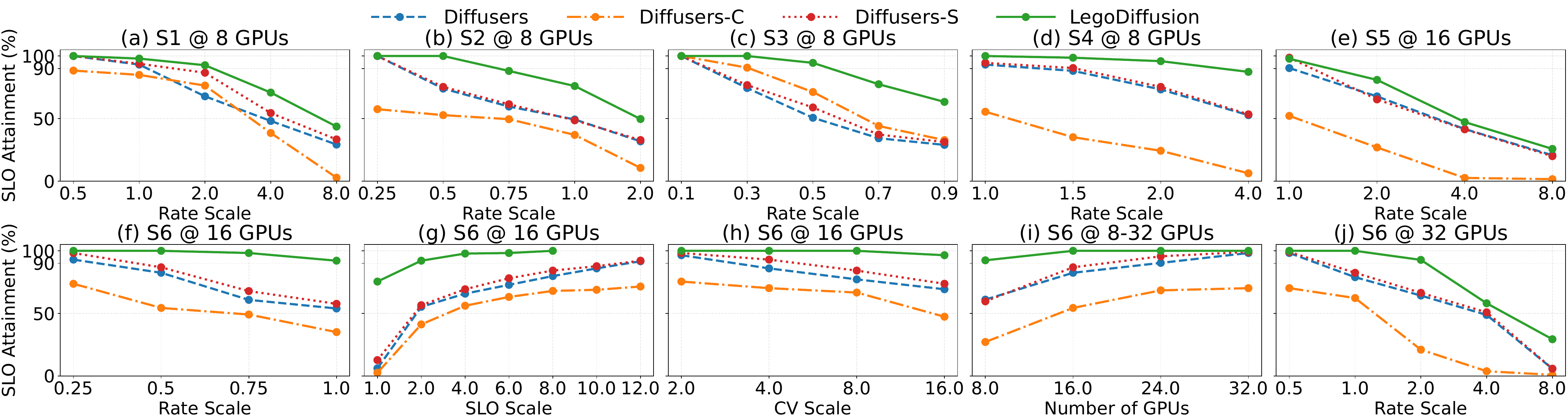}
  \caption{End-to-end performance across six settings (S1--S6 in \tabref{tab:eval_settings}), evaluated under varying request traffic rates (a--f, j), SLO requirements (g), traffic burstiness (h), and testbed sizes (i).}
  \label{fig:e2e_results}
  \vspace{-.2in}
\end{figure*}

As \figref{fig:e2e_results} shows, \sys{} consistently outperforms all baselines
across six settings, four traffic dimensions, and a range of testbed sizes.
At high request rates, \sys{} achieves over 90\% SLO attainment in settings where the strongest baseline drops below 3\%.

\PHM{SLO Attainment vs. Rate.} We evaluate \sys{} and the baselines by varying
the request rate while keeping the SLO scale (2.0) and traffic burstiness fixed.
As shown in \figref{fig:e2e_results} (a)--(f) and (j), \sys{} consistently
achieves higher SLO attainment across varying rate scales. Compared to
\shepherd{}, the strongest baseline, \sys{} sustains up to a 3$\times$ higher
request rate while meeting a 90\% SLO attainment target. At low request rates,
all systems achieve high SLO attainment; however, as the rate increases,
baseline performance plunges due to coarse-grained workflow scaling and
inability to share common models (\S\ref{sec:limitations}). In contrast,
\sys{}'s gains stem from two mechanisms: model sharing
(\S\ref{sec:sched_batching}) enables batching nodes from all three
workflows onto shared model replicas, avoiding redundant loading; adaptive
parallelism (\S\ref{sec:sched_parallelism}) further reduces per-request latency
at low-to-moderate rates by distributing inference across idle GPUs.

Next, we present evaluations that vary the SLO scale, traffic burstiness, and
testbed size, respectively. We focus on the Flux model
family (S6), widely adopted models~\cite{huggingface_model_likes}
representative of recent advances in the field.

\PHM{SLO Attainment vs. SLO Scale (16 GPUs).} In \figref{fig:e2e_results}(g), we fix the
rate scale at 1.0 and evaluate using the original production trace.
Even at a strict SLO scale of 1.0, \sys{} achieves substantially higher SLO
attainment than the baselines.
At this scale, the deadline is tight enough that only intra-node parallelism---splitting the base model across two GPUs (\S\ref{sec:sched_parallelism})---can bring per-request latency below the target; baselines, which run each workflow on a single GPU, cannot meet this deadline regardless of scheduling policy.
At an SLO scale of 2.0, \sys{} satisfies the SLO for over 90\% of requests, whereas the
baselines require an SLO scale of 12.0 to reach the same level. We observe
a sharp increase in baseline SLO attainment when the SLO scale rises from 1.0 to
2.0, because this relaxation begins to absorb the inherent overheads of
monolithic serving (\S\ref{sec:limitations}). Beyond that point, baseline improvements are 
gradual. In contrast,
\sys{} benefits more effectively from relaxed SLOs, achieving \(1.4\times\)
higher SLO attainment than the strongest baseline at an SLO scale of 4.0.

\PHM{SLO Attainment vs. CV (16 GPUs).} In \figref{fig:e2e_results} (h), we fix
the rate scale at 0.25 and the SLO scale at 2.0. Following prior
work~\cite{alpaserve, gujarati2020Clockwork}, we slice the original trace into
time windows and fit the arrivals to a Gamma Process parameterized by the
coefficient of variation (CV). Scaling the CV and resampling allows us to
control traffic burstiness. Higher CVs indicate burstier traffic, which
exacerbates queuing delays and increases SLO violations. As shown, \sys{}
gracefully handles highly bursty traffic, sustaining high attainment even at an
8$\times$ larger CV compared to the baselines. When traffic subsides, the
work-conserving parallelism heuristic (\S\ref{sec:sched_parallelism}) drains the
queue faster by assigning more GPUs per request, creating headroom before the
next burst. Admission control (\S\ref{sec:sched_admission}) then protects
admitted requests during the spike itself by rejecting those that would violate
their SLOs.

\PHM{SLO Attainment vs. Testbed Size.} Finally, in \figref{fig:e2e_results} (i),
we fix the rate scale at 0.5 and the SLO scale at 2.0, varying the testbed size
under the original production trace. \sys{} requires up to 3$\times$ fewer GPUs
to achieve a 90\% SLO attainment target. 
This high resource efficiency is driven by our \emph{micro-serving} design. 
Unlike monolithic serving systems that rigidly partition resources at the workflow level, \sys{} enables fine-grained model scaling and serving, as well as model sharing, which effectively treats all GPUs as a unified pool. 
This eliminates resource over-provisioning, and ensures every available GPU cycle is utilized efficiently.

\begin{figure}[t]
  \centering
  \includegraphics[width=0.49\linewidth]
{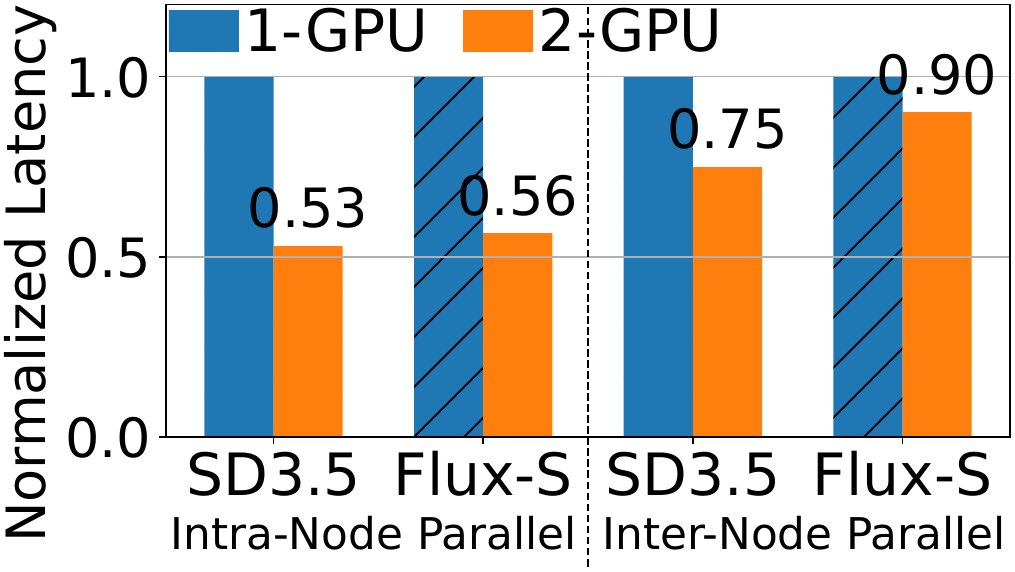}
\includegraphics[width=0.49\linewidth]{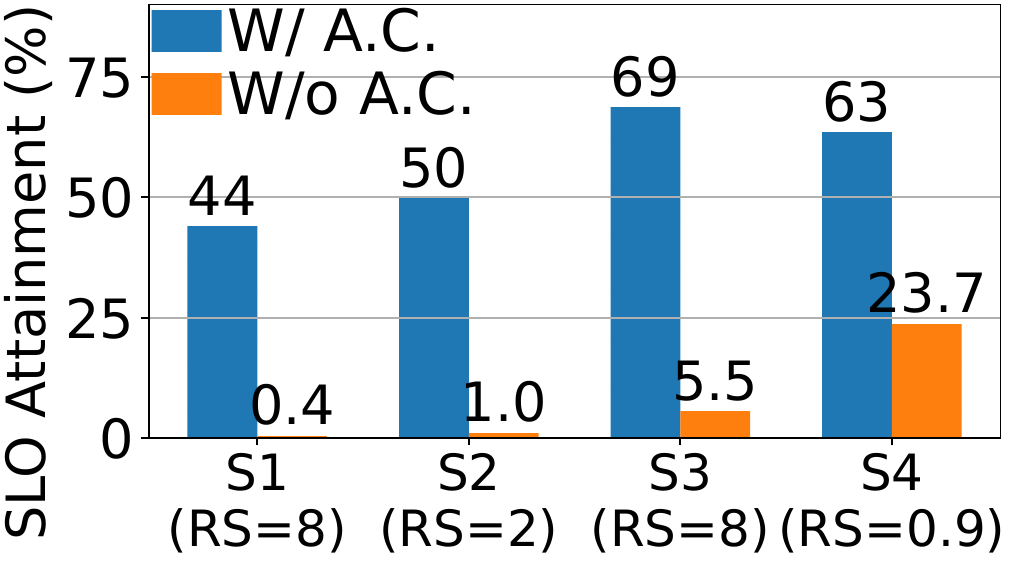}
  \caption{\textbf{Left:} Normalized latency of \sys{} across different numbers of available GPUs with intra-/inter-node parallelism (\S\ref{sec:cluster_optimizations}).  \textbf{Flux-S:} Flux-Schnell. \textbf{Right:} Effectiveness of admission control (A.C.) in settings S1-4. \textbf{RS}: Rate Scale.}
  \label{fig:eval_latent_controlnet_parallel}
  \vspace{-.2in}
\end{figure}

\begin{figure}[t]
  \centering
  \includegraphics[width=1.0\linewidth]
{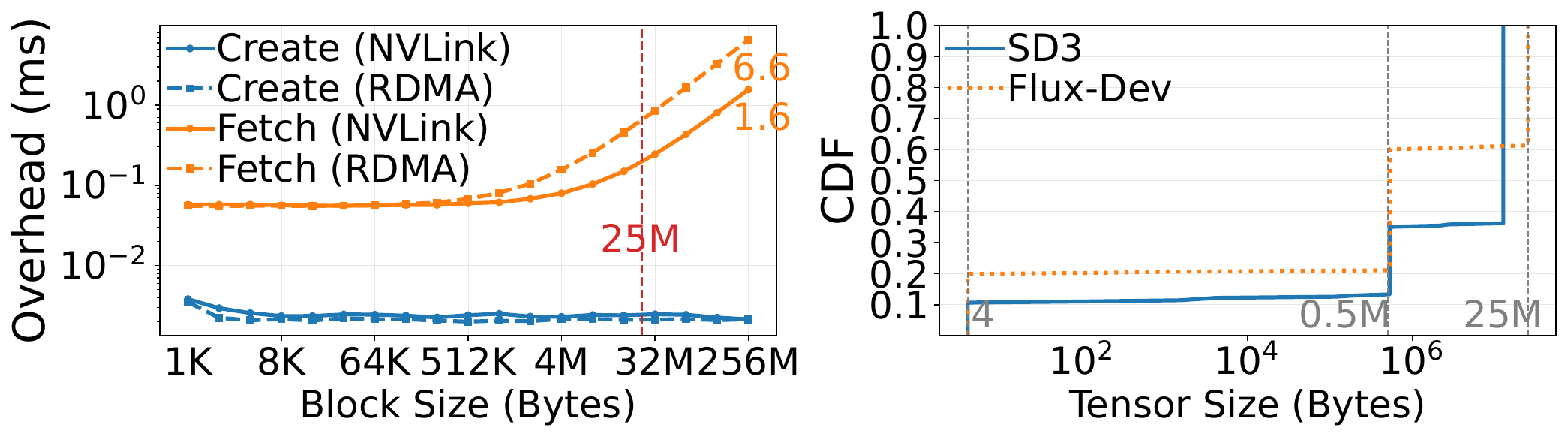}
  \caption{\textbf{Left:} Data fetching latency of varying sizes of tensor blocks. \textbf{Right:} The distribution of tensor block sizes found in typical SD3 and Flux workflows. }
  \label{fig:communication_tensor_size}
  \vspace{-.2in}
\end{figure}

\subsection{Microbenchmarks}
\label{sec:microbenchmark}

We isolate the benefits of \sys{}’s micro-serving (\S\ref{sec:case_for_micro_serving}).

\PHB{Model Sharing.} Diffusion models patched with LoRAs can be shared
across requests, significantly reducing the memory
footprint and latency overhead compared to loading a new model. We validate this
using SD3 and a typical LoRA~\cite{Yarn_art_SD3_LoRA}. While the LoRA occupies
886 MiB of memory and takes 100 ms for swapping~\cite{katz}, this saves the
3.9 GiB of memory and 430 ms of latency incurred by loading a fresh SD3 model.

\PHB{Intra-Node Parallelism.} \sys{} natively integrates latent
parallelism~\cite{li2024DistriFusion, katz, fang2024xdit} with intra-node
parallelism (\S\ref{sec:sched_parallelism}), enabling accelerated diffusion
model inference across two GPUs. As shown in
\figref{fig:eval_latent_controlnet_parallel}-left, our intra-node implementation
achieves a speedup of up to 1.9$\times$. This aligns with findings
in~\cite{katz, li2024DistriFusion} and validates \sys{}'s capability to support
state-of-the-art optimizations.

\PHB{Inter-Node Parallelism.} As
discussed in \S\ref{sec:data_engine}, \sys{} implements a \emph{deferred} data
fetch mechanism to enable ControlNet parallelization~\cite{katz}, a key form of
inter-node parallelism described in \S\ref{sec:sched_parallelism}. As shown
in \figref{fig:eval_latent_controlnet_parallel}-left, \sys{}'s inter-node
parallelism accelerates workflow execution across different models by up to
1.3$\times$, consistent with results in~\cite{katz}. Note that the gains with
Flux models are limited because their ControlNets are small (only 6\% of the
base model size) and have negligible latency compared to the base model.

\PHB{Admission Control.} We evaluate the effectiveness of \sys{}'s admission
control (\S\ref{sec:sched_admission}) in optimizing SLO attainment. In
\figref{fig:eval_latent_controlnet_parallel}-right, enabling admission control
across the four settings (\tabref{tab:eval_settings}) prevents system overload
under high request rates. By proactively aborting requests that are destined to
violate SLOs, \sys{} increases SLO attainment from a mere 0.4\% to 44\% in
setting S1.

\PHB{Programmability.} \sys{} provides an intuitive programming model for
composing complex workflows. Following~\cite{zheng2024sglang}, we quantify
developer productivity using effective Lines of Code (LoC). We compare \sys{}
against Katz~\cite{katz} and xDiT~\cite{fang2024xdit}, two popular diffusion
model serving engines that support parallel acceleration. As shown in
\tabref{tab:productivity}, \sys{} requires comparable or lower implementation
effort to express these optimizations, while additionally supporting adaptive
runtime behavior that the baselines do not.

\begin{table}[t]
\centering
\footnotesize
\renewcommand{\arraystretch}{0.9}
\caption{Effective LOC and adaptive-runtime support.}
\label{tab:productivity}

\begin{tabular}{lccc}
\toprule
\multirow{2}{*}{\textbf{Technique}} 
& \multicolumn{3}{c}{\textbf{LOC / (Support adaptive adjustment?)}} \\
& \textbf{Katz}~\cite{katz} 
& \textbf{xDiT}~\cite{fang2024xdit} 
& \textbf{\sys{}} \\ 
\cmidrule(l){1-4}

Latent parallel     & 92 (No) & 68 (No) & 74 (Yes) \\
ControlNet parallel & 127 (No) & N.A. & 79 (Yes) \\
Async LoRA loading  & 182 (Yes) & N.A. & 61 (Yes) \\ 

\bottomrule
\end{tabular}
\vspace{-0.2in}
\end{table}

\subsection{Case Study}
\label{sec:eval_case_study}

We next show that \sys{} supports emerging optimizations
tailored for diffusion models described in
\S\ref{sec:compiler_with_optimizations}.

\PHB{Approximate Caching.} 
In \sys{}, we implement Nirvana's~\cite{nirvana} \emph{approximate caching}
optimization (\S\ref{sec:diffusion_primer}). Following prior work~\cite{nirvana,
katz}, we use a SDXL workflow and configure to reduce 20\% and 40\% denoising
computation, respectively. The optimization achieves speedups of 1.13$\times$
and 1.43$\times$ with its original implementation on \diffusers{}, and
comparable speedups of 1.17$\times$ and 1.42$\times$ on \sys{}, evidencing
\sys{}'s effective support for the optimization.

\PHB{Async LoRA Loading.} We implement Katz's~\cite{katz} asynchronous LoRA
loading design in \sys{} and evaluate it against the original \diffusers{}
implementation, using SDXL~\cite{podell2024sdxl} with a papercut-style
LoRA~\cite{papercut_lora}. Our implementation reduces LoRA loading overhead from
0.5 seconds to 0.05 seconds, matching the results reported in~\cite{katz}.

\subsection{System Overhead}
\label{sec:system_overhead}

\PHB{Execution Overhead.} Micro-serving introduces additional
overhead due to inter-node communication and control-plane coordination.
We quantify this overhead by comparing \sys{} against monolithic baselines on
four workflows: SD3, SD3.5-Large, Flux-Dev, and Flux-Schnell. Across all cases,
the maximum end-to-end overhead is 150\,ms. Given that these diffusion workloads
typically take 2--20 seconds to complete, this additional cost is negligible.

\PHM{Control-Plane Scalability.} To show \sys{}'s control plane remains
efficient at large scale, we conduct simulation-based experiments on a 256-GPU
setup under high concurrency, with 500 inflight requests. Across two
representative workloads, Flux-Dev and SD3.5-Large, the coordinator accounts for
only 3.4\% and 2.7\% of total execution time, respectively, indicating that the
control plane does not emerge as the dominant bottleneck at this scale.

\PHM{Data Transmission Latency.} We further isolate the cost of intermediate
tensor movement in \figref{fig:communication_tensor_size}. The left panel shows
the latency of tensor serialization and transmission over a range of tensor
sizes, and the right panel reports the actual intermediate tensor sizes produced
by SD3 and Flux-Dev workflows with ControlNet. Even for the largest intermediate
tensors, transmission latency remains below 1\,ms,
confirming that the inter-GPU bandwidth is not a bottleneck for \sys{}'s
fine-grained execution model.

\section{Discussion and Related Works}
\label{sec:related_works}

\PHB{Scalability and Fault Tolerance.}
\label{sec:scalability_fault_tolerance}
In \sys{}, one coordinator manages $N$ executors (\figref{fig:architecture}).
To avoid a coordinator bottleneck as $N$ grows, \sys{} shards executors across
multiple coordinators, each managing a disjoint subset of workflows that share
models, preserving sharing opportunities. A cluster management
service~\cite{zookeeper,zookeeper_atc} handles coordinator discovery and failure
detection. Executor failures are tolerated naturally: the coordinator reassigns
affected nodes to other executors.

\PHM{Diffusion Model Serving Systems.} Existing serving
systems~\cite{diffusers_server, comfyUI_server, sglang_diffusion, vllm_omni}
follow a monolithic design with limited adapter
support~\cite{sglang_adapter_pr, vllm_omni, sglang_roadmap2025}
(\S\ref{sec:challenges}). Several works accelerate individual workflow
execution: Nirvana~\cite{nirvana} reduces denoising steps via cached images;
DistriFusion~\cite{li2024DistriFusion} and xDiT~\cite{fang2024xdit} exploit
multi-GPU parallelism; Katz~\cite{katz} parallelizes ControlNets and
asynchronously loads LoRAs;
TetriServe~\cite{lu2026tetriserveefficientditserving} and
TridentServe~\cite{xia2025tridentservestagelevelservingdiffusion} adapt sequence
parallelism for latency SLOs. However, several of
these~\cite{xia2025tridentservestagelevelservingdiffusion,
lu2026tetriserveefficientditserving, nirvana, li2024DistriFusion} lack adapter
support prevalent in production~\cite{katz, diffusion_production}, and none
target cluster-level multi-workflow deployment. \sys{}'s micro-serving approach
is complementary (\S\ref{sec:compiler_with_optimizations},
\S\ref{sec:eval_case_study}).

\PHM{Other Model Serving Systems.} Prior work on model serving has improved
latency~\cite{crankshaw2017Clipper, wang2023Tabi},
throughput~\cite{ahmad2024Proteus, yang2022INFless}, and resource
efficiency~\cite{zhang2019MArk, wang2021Morphling, gunasekaran2022Cocktail,
yang2025Prism, wang2023MGG} across DNNs and LLMs~\cite{yu2022Orca, sarathi,
MuxServe, loongserve,helix,HedraRAG,CacheBlend,oliaro2025FlexLLM,
JITServe,ic_cache, LLMStation, DeltaZip, ktransformer, srivatsa2025preble,
Mobile_Soc_llm, Medusa, WEAVER}. \sys{} complements them by focusing on
text-to-image serving, which has distinct computational and workflow
characteristics. Prior work on online multi-model serving has further explored
model placement~\cite{zhang2023SHEPHERD, alpaserve}, request
scheduling~\cite{zhang2023SHEPHERD,gujarati2020Clockwork}, and dynamic
scaling~\cite{serverlessLLM,zhang2024fast}. In \S\ref{sec:evaluation}, we
integrate representative techniques from them with existing diffusion workflow
serving systems~\cite{diffusers_server}. Despite their effectiveness, \sys{}
outperforms with its \emph{micro-serving} approach, which fundamentally
addresses the limitations of \emph{monolithic} serving. Also, \sys{} is
compatible with these optimizations.
\section{Conclusions}
\label{sec:conclusions}
We presented \sys{}, an efficient micro-serving system for diffusion workflow. 
\sys{} has three key designs: (1) a programming model that transforms workflow compositions into loosely coupled nodes; (2) a specialized runtime and data plane that streamline data communication to facilitate micro-serving; and (3) a scheduler that realizes the benefits of micro-serving at the cluster level. Collectively, \sys{} outperforms existing \emph{monolithic} serving systems, sustaining up to 3$\times$ higher request rates and tolerating 8$\times$ higher burst traffic, with the same performance requirements.

\bibliographystyle{ACM-Reference-Format}
\bibliography{main}

\end{document}